\documentclass{article}
\usepackage{graphicx,amssymb}
\usepackage[rightcaption]{sidecap}
\usepackage{natbib}
\usepackage[margin=0.5in]{geometry}
\usepackage{color}
\usepackage{enumitem}
\usepackage{authblk}
\usepackage{hyperref}
\usepackage{threeparttable}
\usepackage{enumitem}
\usepackage[compact]{titlesec}
\titleformat{\section}
  {\normalfont\large\bfseries} 
  {\thesection}
  {1em}
  {}

\title{Multi-epoch imaging observations of the Galactic Center with NIRCam and MIRI}

\author[1]{Rainer Sch{\"o}del}
\author[2]{Adam Ginsburg}
\author[3,4]{Steve Longmore}
\author[1]{Francisco Nogueras-Lara}
\author[5]{John Bally}
\author[6]{Samantha Adams}
\author[1]{Mar\'ia \'{A}lvarez-Santiago}
\author[7,8,9]{Pau Amaro Seoane}
\author[10]{Konstantina Anastasopoulou}
\author[11,12]{Mayura Balakrishnan}
\author[13]{A.~T.~Barnes}
\author[14]{Giuseppe Bono}
\author[1]{Luc\'ia Bravo Ferres}
\author[15]{Aaron Bryant}
\author[2]{Nazar Budaiev}
\author[16,17]{Santi Cassisi}
\author[18]{Anna Ciurlo}
\author[19]{Angela S. Cotera}
\author[18]{Tuan Do}
\author[20]{Marco Donati}
\author[8]{Frank Eisenhauer}
\author[1]{Rub\'en Fedriani}
\author[21]{Naomi Fee Schm\"okel}
\author[22]{Anja Feldmeier-Krause}
\author[20,23]{Karl Fiteni}
\author[11,12]{Nicole M. Ford}
\author[1]{A. Teresa Gallego Calvente}
\author[24]{Macarena Garcia Marin}
\author[25,26]{Pablo Garc\'ia}
\author[1]{Angela Gardini}
\author[18]{Andrea M. Ghez}
\author[2]{Savannah R. Gramze}
\author[21]{Lena Gro\ss{}ekath\"ofer}
\author[27]{Alessia Gualandris}
\author[10]{Mario G. Guarcello}
\author[11,12]{Daryl Haggard}
\author[28]{Matthew  J.\ Hankins}
\author[29]{Jonathan~D.~Henshaw}
\author[18]{Matthew W. Hosek Jr.}
\author[30]{Yue Hu}
\author[31]{Namitha Issac}
\author[32,33]{Ralf S.\ Klessen}
\author[4]{J.\ M.\ Diederik Kruijssen}
\author[34]{Alex Lazarian}
\author[20]{Xingchen Li}
\author[35]{Mattia Libralato}
\author[36]{Dani R. Lipman}
\author[37]{Xing Lu}
\author[38]{Shifra Mandel}
\author[39]{Roeland P. van der Marel}
\author[41,42,43]{Sera Markoff}
\author[1]{\'{A}lvaro Mart\'inez-Arranz}
\author[35,44,45]{Alessandra Mastrobuono-Battisti}
\author[21]{Maria Melamed}
\author[46]{Joseph M. Michail}
\author[38]{Kaya Mori}
\author[18]{Mark R. Morris}
\author[47,48,49]{Lena Murchikova} 
\author[50]{Francisco Najarro}
\author[51]{David Nataf}
\author[29]{Nadine Neumayer}
\author[52]{Shogo Nishiyama}
\author[53]{Masayoshi Nobukawa}
\author[54,55]{Xing Pan}
\author[56]{Maya A.\ Petkova}
\author[57,8,20]{Gabriele Ponti}
\author[58]{Pau Ramos}
\author[59]{Carlos G. Rom\'an-Z\'u\~niga}
\author[60]{Nils Ryde}
\author[61]{Nadeen B. Sabha}
\author[62]{Claudio Salvo-Guajardo}
\author[63,64]{\'Alvaro S\'anchez-Monge}
\author[68]{Mathias Schultheis}
\author[65]{Stephen Serjeant}
\author[9]{Lijing Shao}
\author[19]{Janet P. Simpson}
\author[20]{Mattia~C.~Sormani}
\author[66]{Shunya Takekawa}
\author[60,68]{Brian Thorsbro}
\author[69]{Pablo Torne}
\author[29]{Robin Tress}
\author[62]{\'Alvaro Valenzuela~Navarro}
\author[70]{Daniel L. Walker}
\author[29]{Fengwei Xu}
\author[2]{Siyao Xu}
\author[71]{Michal Zaja\v{c}ek}
\author[46]{Qizhou Zhang}
\author[62]{Manuela Zoccali}

\affil[1]{Instituto de Astrof\'isica de Andaluc\'ia (CSIC), Granada, Spain}
\affil[2]{University of Florida, Gainesville, FL, USA}
\affil[3]{Liverpool John Moores University, Liverpool, UK}
\affil[4]{Cosmic Origins Of Life (COOL) Research DAO, \href{coolresearch.io}{coolresearch.io}`}
\affil[5]{University of Colorado Boulder, Boulder, CO, USA}
\affil[6]{Department of Physics, University of Connecticut, CT, USA}
\affil[7]{Universitat Polit\`ecnica de Val\`encia, Spain}
\affil[8]{Max Planck Institute for extraterrestrial Physics, Garching, Germany}
\affil[9]{Kavli Institute for Astronomy and Astrophysics, Peking University, Beijing, China}
\affil[10]{INAF -- Osservatorio Astronomico di Palermo,  Palermo, Italy}
\affil[11]{Department of Physics, McGill University, Qu\'ebec, , Canada}
\affil[12]{Trottier Space Institute at McGill, Montr\'eal, Qu\'ebec, Canada}
\affil[13]{European Southern Observatory (ESO), Garching, Germany}
\affil[14]{University of Rome Tor Vergata, Italy }
\affil[15]{Institut f{\"u}r Raumfahrtsysteme, Universit{\"a}t Stuttgart, Stuttgart, Germany}
\affil[16]{INAF - Astronomical Observatory of Abruzzo, Teramo, Italy}
\affil[17]{INFN, National Institute for Nuclear Physics, Section of Pisa, Pisa, Italy}
\affil[18]{University of California Los Angeles, Physics and Astronomy Department, Los Angeles, USA}
\affil[19]{SETI Institute,Mountain View, CA, USA}
\affil[20]{Como Lake centre for AstroPhysics (CLAP), DiSAT, Universit{\`a} dell'Insubria, Como, Italy}
\affil[21]{I. Physikalisches Institut, Universit\"at zu K\"oln, K\"oln, Germany}
\affil[22]{University of Vienna, Vienna, Austria}
\affil[23]{Institute of Space Sciences \& Astronomy, University of Malta, Msida, Malta}
\affil[24]{European Space Agency, ESA Office, Space Telescope Science Institute,  Baltimore, MD, USA}
\affil[25]{Chinese Academy of Sciences South America Center for Astronomy, National Astronomical Observatories, CAS, Beijing, China}
\affil[26]{Instituto de Astronom\'ia, Universidad Cat\'olica del Norte, Antofagasta, Chile}
\affil[27]{Department of Mathematics and Physics, University of Surrey, Guildford, United Kingdom}
\affil[28]{Arkansas Tech University, Russellville, AR, USA}
\affil[29]{Max-Planck-Institut f\"ur Astronomie, Heidelberg, Germany}
\affil[30]{California Institute of Technology, Pasadena, CA, USA}
\affil[31]{Korea Astronomy and Space Science Institute, Daejeon, Republic of Korea}
\affil[32]{Universit\"{a}t Heidelberg, Zentrum f\"{u}r Astronomie, Institut f\"{u}r Theoretische Astrophysik, Heidelberg, Germany}
\affil[33]{Universit\"{a}t Heidelberg, Interdisziplin\"{a}res Zentrum f\"{u}r Wissenschaftliches Rechnen, Heidelberg, Germany}
\affil[34]{University of Wisconsin, Madison, USA}
\affil[35]{INAF - Osservatorio Astronomico di Padova, Padova, Italy}
\affil[36]{Department of Physics, University of Connecticut, Storrs, CT, USA}
\affil[37]{Shanghai Astronomical Observatory, Chinese Academy of Sciences, Shanghai, People's Republic of China}
\affil[38]{Columbia Astrophysics Laboratory, Columbia University, New York, NY, USA}
\affil[39]{Space Telescope Science Institute,Baltimore, MD, USA}
\affil[40]{Center for Astrophysical Sciences, The William H. Miller III Department of Physics \& Astronomy, Johns Hopkins University, Baltimore, MD, USA}
\affil[41]{Institute of Astronomy, University of Cambridge, Madingley Road, CB3 0HA, United Kingdom}
\affil[42]{Anton Pannekoek Institute for Astronomy, University of Amsterdam, Amsterdam, The Netherlands}
\affil[43]{Gravitation and Astroparticle Physics Amsterdam (GRAPPA) Institute, University of Amsterdam, Amsterdam, The Netherlands}
\affil[44]{Dipartimento di Fisica e Astronomia ``Galileo Galilei'', Universit\`a di Padova,  Padova, Italy}
\affil[45]{Dipartimento di Tecnica e Gestione dei Sistemi Industriali, Universit\`a di Padova, Vicenza, Italy}
\affil[46]{Center for Astrophysics $\vert$ Harvard \& Smithsonian, Cambridge, MA, USA}
\affil[47]{Center for Interdisciplinary Exploration and Research in Astronomy, Northwestern University, Evanston, IL, USA}
\affil[48]{Department of Physics \& Astronomy, Northwestern University, Evanston, IL, USA}
\affil[49]{School of Natural Sciences, Institute for Advanced Study, Princeton, NJ, USA}
\affil[50]{Centro de Astrobiolog\'{i}a, CSIC-INTA, Torrejon de Ardoz, Spain}
\affil[51]{University of Iowa, Iowa, USA }
\affil[52]{Miyagi University of Education, Sendai, Miyagi, Japan}
\affil[53]{Nara University of Education, Nara, Japan}
\affil[54]{School of Astronomy and Space Science, Nanjing University, Nanjing, P.R.China}
\affil[55]{Key Laboratory of Modern Astronomy and Astrophysics (Nanjing University), Ministry of Education, Nanjing , P.R.China}
\affil[56]{Department of Physics and Astronomy, Chalmers University of Technology, Gothenburg, Sweden}
\affil[57]{INAF -- Osservatorio Astronomico di Brera, Merate, Italy}
\affil[58]{National Astronomical Observatory of Japan, JASMINE project, Tokyo, Japan}
\affil[59]{Instituto de Astronom\'ia, Universidada Nacional Aut\'onoma de M\'exico, Ensenada Mexico}
\affil[60]{Division of Astrophysics, Department of Physics, Lund University, Lund, Sweden}
\affil[61]{Universit\"at Innsbruck, Institut f\"ur Astro- und Teilchenphysik, Innsbruck, Austria}
\affil[62]{Instituto de Astrof\'isica, Pontificia Universidad Cat\'olica de Chile, Santiago, Chile}
\affil[63]{Institut de Ci\`encies de l'Espai (ICE), CSIC, Barcelona, Spain}
\affil[64]{Institut d'Estudis Espacials de Catalunya (IEEC), Barcelona, Spain}
\affil[65]{The Open University, School of Physical Sciences, Walton Hall, Milton Keynes, UK}
\affil[66]{Faculty of Engineering, Kanagawa University, Yokohama, Japan}
\affil[67]{Dept. Of Astronomy, University of Virginia, Charlottesville, VA, USA}
\affil[68]{Universit\'e C\^ote d'Azur, Observatoire de la C\^ote d'Azur,
CNRS, Nice, France}
\affil[69]{Institut de Radioastronomie Millim\'etrique (IRAM), Granada, Spain}
\affil[70]{UK ALMA Regional Centre Node, Jodrell Bank Centre for Astrophysics, The University of Manchester, Manchester, UK}
\affil[71]{Department of Theoretical Physics and Astrophysics, Faculty of Science, Masaryk University, Brno, Czechia}

\date{\today}

\begin{document}

\maketitle

\newpage

\section*{Abstract}

We propose a multi-cycle Treasury Program of NIRCam and MIRI imaging of the nuclear stellar disk/central molecular zone of the Milky Way. The estimated required observing time is 420\,h, distributed over three cycles spaced about four years apart.

\section{Introduction}

The center of the Milky Way is the densest and most extreme environment in our Galaxy, its most active star forming region, and the only galaxy nucleus where we can resolve stars down to about a solar mass on the main sequence. The Galactic Center (GC) contains a nuclear stellar disk, a nuclear star cluster, and a supermassive black hole, all embedded within the so-called Central Molecular Zone, which showcases conditions of the interstellar medium that are typically found in high-redshift star-forming regions. The GC  hosts  the highest concentration of massive stars in the Milky Way and hosts a large concentration of rare objects, such as Wolf Rayet stars and X-ray binaries. The GC is therefore an astrophysical target of outstanding  importance.

A multi-epoch JWST program enables novel science including the lifetime of planet-forming disks,  the search for intermediate mass black holes, direct measurements of the multi-scale gravitational potential, and a deep measurement of the star formation history.
Multi-epoch imaging observations with JWST's NIRCam and MIRI instruments will go far beyond Roman WFI surveys  in terms of depth and astrometric and photometric precision because of its $3\times$ better resolution, and it expands spectral coverage in narrow bands and mid-infrared wavelength coverage. 
The proposed observations would correspond to a Treasury Program of about 420\,h, distributed over three epochs spaced three to five years apart.

\section{The Galactic Center: a target of fundamental astrophysical importance}
\label{sec:goals}

The astrophysical community has identified and elaborated on several major open science questions that can be addressed by studying the GC with JWST \citep[see][]{Schoedel:2023rf}: 

\begin{enumerate}[nosep,label=(\Alph*)]
        \item What is the formation history of the GC and its relation to the overall formation history of the Milky Way?
        \item Is the overall energetics of the GC dominated by episodic AGN activity or feedback from star formation?
        \item What is the origin of, and environmental variation in, the stellar initial mass function?
        \item Why is the star formation rate an order of magnitude lower than predicted by star-formation-dense-gas relations?
        \item What is the 3D structure of the interstellar medium (ISM) orbiting and fuelling accretion and star formation?
\end{enumerate}

\begin{figure}[!b]
\centering
\includegraphics[width=.8\textwidth]{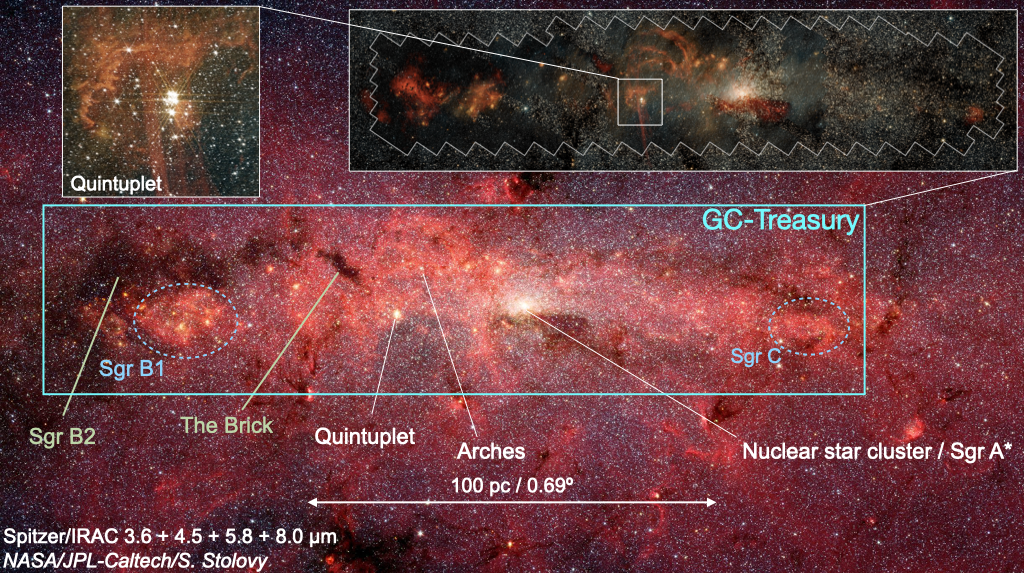}
\caption{Overview of the Galactic Center. Galactic north is up and east to the left. Prominent H\,II regions are labelled in blue, star clusters in white, and major molecular clouds in green. The large rectangle outlines the GC-Treasury survey area. The upper right inset shows an overview of the JWST GC Treasury Survey with the white polygon indicating the exact field.  VVV Ks $+$ Spitzer $4.5\,\mu$m  form the backdrop, with JWST F212N $+$ F480M from the Sep.\,2026 campaign on top.  ALMA ACES 3\,mm continuum in red shows ionized gas and dust (\href{https://starformation.astro.ufl.edu/avm_images/GCSpitzerLike_vvv_aces_hips/}{An interactive full resolution image can be inspected online}). The upper left inset shows a zoom onto the Quintuplet, as seen with GC-Treasury. \label{fig:overview}}
\end{figure}

\section{The JWST GC Treasury Survey}

The JWST Galactic Center Treasury Survey \href{https://www.stsci.edu/jwst-program-info/program/?program=10678}{GO 10678} (GC-Treasury hereafter), is a first, groundbreaking step to address these questions with the unique capabilities of the JWST. GC-Treasury images the area outlined by the Central Molecular Zone (CMZ) and Nuclear Stellar disk (NSD) with the short F212N and long F480M NIRCam filters and half of the area with MIRI F770W (Fig.\,\ref{fig:overview}). Half of the data were acquired in Sep. 2026, the remaining observations are scheduled for Sep. 2027.

\href{https://sites.google.com/view/jwst-gc/home}{JWST GC-Treasury} is an open, foundational project that will allow the community to obtain a clear view of the size and structure of the nuclear star cluster and nuclear stellar disk, constrain the star formation history of the GC via luminosity functions, find candidates for so far unknown massive young clusters and stars, measure the faint end of the initial mass function of the Arches and Quintuplet young massive clusters, constrain the current star formation activity via finding YSO candidates, and substantially improve our knowledge of the 3D distribution of molecular clouds at the GC.

Deep, sensitive observations of the extremely extinguished GC ($A_{K}\approx2-10$\,mag) require going to long wavelengths, on the one hand, and high angular resolution, on the other hand. This limits precise, accurate Roman measurements to stars brighter than $[F213]\approx18$, above the main sequence turn-off of the mostly old stellar population in the GC. The impact of crowding on faint stars is illustrated in Fig.\,\ref{fig:comparison}.
JWST, by contrast, can reach at least down to $F212N\approx$21\,mag, recovering $\sim10\times$ more stars.

\section{The fundamental impact of multi-epoch NIRCam and MIRI observations}
\label{sec:multiepoch}

We have organised the case for the
multi-epoch survey around ten science questions (Q1--Q10) that GC-Treasury alone and the planned Roman surveys cannot answer. Table~\ref{tab:epochs} lists the data and science capabilities that become
available after each epoch and marks the point at which each question can be answered.

Each new epoch of JWST GC coverage builds on the previous ones by extending the time baseline for proper motion measurements and adding filters for measurements of star and ISM properties. Multiple epochs will be essential to cover the millions of sources Roman will not measure and enable proper acceleration measurements once four epochs are completed.
Each epoch adds 2.5 additional filter bands, two from NIRCam and half from MIRI, for a final target of 10 filter coverage, producing rich spectral energy distribution measurements capable of simultaneously measuring both the stars and absorbing ISM.
The extended emission maps support ISM measurements that ground extragalactic observations.

The astrometric uncertainties in Table~\ref{tab:epochs} assume a positional uncertainty of $\sigma = 0.5$\,mas per epoch for stars with $F212N \lesssim 18$\,mag, and an interval of four years between epochs. Proper motion uncertainties scale as $\sigma/T$ for a total baseline $T$, and acceleration uncertainties scale as $\sigma/\Delta t^{2}$ for an interval $\Delta t$ between epochs. 
A total time baseline of 12 years would therefore give a proper motion uncertainty of $0.04$\,mas\,yr$^{-1}$ and, assuming four year spacing between epochs, an acceleration uncertainty 
0.031\,mas\,yr$^{-2}$. At a distance of 8.2\,kpc, 1\,mas\,yr$^{-1}$ corresponds to
39\,km\,s$^{-1}$. 

With JWST we expect to measure the proper motions of about ten times as many stars as with Roman (see right panel of Fig.\,\ref{fig:comparison}), and can also unambiguously cross-match sources brighter than  $F212N \approx 21$\,mag across epochs and filters. 

Narrow-band imaging (Pa$\alpha$ and Br$\alpha$, H$_2$) traces shocks arising from sources such as from supernovae, stellar winds, proto-stellar outflows, and planetary nebulae and will probe feedback impacts on the CMZ gas. HST Pa$\alpha$ observations have demonstrated the power of this method \citep{Wang:2010fk}.  JWST Br$\alpha$ imaging of  Sgr\,C
have already revealed remarkable filamentary morphology suspected to be produced by strong magnetic fields \citep{Bally:2024iu}.
Narrowband measurements of stars reveal accretion, either from ISM onto YSOs or mass transfer from stars to compact objects.

\subsection{Q1. Where are the missing young stars and clusters?}

Integrated over the past $\sim$30\,Myr, the present-day star formation rate of the CMZ of $\sim$0.1\,$M_{\odot}$\,yr$^{-1}$ \citep{Longmore:2013yu,Barnes:2017gl,Hatchfield:2024zk} would have
produced a few $10^{6}\,M_{\odot}$ of stars \citep{Nogueras-Lara:2020pp}. For a long time only the Arches, Quintuplet and central parsec massive young clusters have been known, but they contain at most a tenth of this mass. There is evidence \citep{Ginsburg:2018oz} that $>1/3$ of stars in the GC form in clusters, which implies the existence of so far unknown, diluted or hidden, clusters in the GC.  In the extremely dense and reddened GC environment, missing clusters can only be found using a combination of colors and proper motions, and only JWST has the depth to identify enough co-moving stars. Although clusters disrupt within a few Myrs  \citep{Kim:2000qt,Portegies-Zwart:2002fk,Kruijssen:2014ok}, dissolving clusters remain kinematically coherent long after they have stopped being over-densities.  \citet{Martinez-Arranz:2024gm} have identified several such groups in a pilot study and confirmed one of them as a young  cluster, Candela 1 \citep{Martinez-Arranz:2024nr}. Three-epoch proper motions with uncertainties of $\sim$3\,km\,s$^{-1}$, comparable to the internal velocity dispersion of a young cluster, would be ideal to identify the members of co-moving groups with high probability. Because proper motions also separate the far side of the nuclear stellar disk from the near side, the same data allow a search for clusters hidden behind the
densest gas (see Q4).

\subsection{Q2. Is the IMF top-heavy?}
The initial mass function (IMF) is of fundamental importance for most fields of astrophysics. Current observational evidence suggests that it is top-heavy at the GC, producing a significantly higher fraction of high-mass stars than the Milky Way's disk. However, this is only based on the Arches, Quintuplet, and central parsec clusters \citep{Bartko:2010fk,Hosek:2019vn,Rui:2019ch,Gallego-Cano:2024lm,Cano-Gonzalez:2025ay}. Identifying the missing clusters (Q1) and non-cluster-formed young stars (Q5) and characterizing them via multi-filter photometry is essential for determining whether the IMF variation is a global property of the GC.

\subsection{Q3. How do the properties of dust and ice vary with environment?}

With one color per star, the amount of extinction and the shape of the extinction
curve are degenerate. At the third epoch every star has photometry in six bands between
1.4 and 4.8\,$\mu$m, enough to fit the extinction curve star by star and to map how its
shape changes across the CMZ. F335M covers the 3.3\,$\mu$m aromatic emission band, which
together with the complete MIRI F770W map of the 7.7\,$\mu$m band traces PAH emission
across the region. The fourth epoch adds F466N, which measures CO ice absorption
towards background stars against the continuum defined by F405N and F480M, and it
completes the F2100W map of warm dust emission. 
For this final epoch filter, it is possible that another option (e.g., F430M, sensitive to CO$_2$) is better for anchoring the ice component of the extinction curve; ongoing spectroscopic studies will decide this.

\subsection{Q4. What is the three-dimensional structure of the dense gas in the CMZ?}

The arrangement of the CMZ clouds along the line of sight is still debated, and competing models place some of the same clouds on opposite sides of Sgr~A$^{*}$ \citep{Henshaw:2022nm}.
Stars provide an independent test. Because the NSD rotates, stars on its near and far sides move in opposite directions in Galactic longitude \citep{Schonrich:2015uq,Shahzamanian:2022vz}, so proper motions place stars statistically in front of or behind the mid-plane, independently of their
extinction \citep{Martinez-Arranz:2022wj,Nogueras-Lara:2026ez,Vasini:2026eb}. Combining this with the star-by-star extinctions from the multi-band photometry locates each cloud along the line of sight through the extinction step between the stars in front of it and those behind it. Two epochs are enough to separate the near- and far-side populations among the bright stars, and three epochs give
reliable motions for the much larger number of faint stars needed to map individual clouds at high angular resolution.

\subsection{Q5. What is the present-day star formation rate?}

The youngest, still accreting YSOs are deeply embedded and require long-wavelength data to be detected and characterized.
While the single-epoch treasury program supplies a rich catalog of YSO candidates, a single epoch and a single IR excess band cannot measure source properties, particularly because of ambiguities between AGB stars and YSOs \citep{Koepferl:2015aq,Nandakumar:2018rw}.
For the youngest objects, the complete MIRI F770W map after the second epoch and the F2100W map after the fourth will provide the spectral coverage needed to break these degeneracies.
Source variability and Br$\alpha$ emission identify accreting objects and constrain their accretion rates. 
The multiepoch NIRCam-long proper motion measurements will identify comoving groups of young stars, tying them to their birth environments.

\subsection{Q6. Can planet-forming disks survive in an environment like that of the early Universe?}

The GC is the only place where we can resolve individual stars of (sub)solar mass forming under the high gas pressures and intense radiation fields that were typical of star formation at redshift 2 to 3 \citep{Kruijssen:2013tq}. Whether disks survive there for long enough to form planets sets the initial conditions for planet formation over most of cosmic history. Answering this needs a census of disks around stars that are confirmed members of young
clusters and associations (Q1, Q5). The known young clusters move at several mas\,yr$^{-1}$ relative to the field population \citep{Hosek:2019vn, Rui:2019ch,Martinez-Arranz:2024gm,Martinez-Arranz:2024nr}, so proper motions from the second epoch already separate their bright members, and the
third epoch extends reliable membership to the low-mass stars that host most disks. Pa$\alpha$ (F187N) and Br$\alpha$ (F405N) excess emission traces ongoing accretion
\citep{Alcala:2017fe}, and  photometry from 1.4 to 4.8\,$\mu$m measures the infrared excess from inner disks \citep{Stolte:2015vn}. Disk fractions can then be measured as a function of cluster age and of distance from the massive stars that drive external photoevaporation.
Separating reddening from excess emission requires at least 5, ideally 10 bands.

\subsection{Q7. Where and how many stellar-mass black holes and X-ray binaries are in the GC?}

We expect that more than $10^{6}$ stellar mass black holes exist in the NSD \citep{Schoedel:2026bf}. If the initial mass function in the GC is top-heavy (Q2; \citealt[e.g.][]{Hosek:2019vn,Gallego-Cano:2024lm}), there may be a factor of a few more. The NSC is also an ideal place to look for high mass merger products of stellar mass black holes \citep{Fragione:2022bd}.
X-ray observations  can only  reveal limited
 information about thousands of sources \citep{Zhu:2018el}, but multi-epoch NIRCam data can identify them from acceleration and recombination line excess measurements. A solar-mass star orbiting a 10\,$M_{\odot}$ black hole with $P=10$ yr has a reflex amplitude of 1.8\,mas, a measurement within reach of the suggested program and definitely possible for brighter stars and/or longer baselines. 
The infrared counterparts of accreting X-ray binaries can be identified through the Pa$\alpha$ and Br$\alpha$ emission from their accretion
flows (combined with by their lack of H$_2$ emission) and through variability.

\subsection{Q8. Is there a population of unseen intermediate-mass black holes in the GC?}

IMBHs may be delivered to the GC by inspiralling star clusters \citep{Bellovary:2025bd}, and several candidates have been proposed \citep{Oka:2016ys,Peissker:2024ly}, but none has been confirmed. Multi-epoch astrometry provides two independent tests. Stars bound within 0.05\,pc (1.2\,arcsec) of a $10^{4}\,M_{\odot}$ IMBH orbit at about 30\,km\,s$^{-1}$ (0.75\,mas\,yr$^{-1}$), nearly ten times the three-epoch proper motion uncertainty, so
after the third epoch a compact group of faint stars with an anomalously large velocity
dispersion would stand out against the field. The fourth epoch measures accelerations
directly. A star 1000\,AU (0.12\,arcsec) from an IMBH of $10^{5}\,M_{\odot}$ accelerates
at 0.5\,mas\,yr$^{-2}$, more than ten times the four-epoch uncertainty, so a single star
would reveal it. For a $10^{4}\,M_{\odot}$ IMBH the acceleration at the same distance
falls to 0.05\,mas\,yr$^{-2}$, which is marginal, and the velocity dispersion test is
then the more sensitive of the two.

\subsection{Q9. How and when did the nuclear stellar disk and nuclear star cluster assemble?}

The nuclear stellar disk (NSD), the nuclear star cluster (NSC) and the inner Galactic
bar overlap along every line of sight towards the GC, so luminosity functions from a
single epoch average over structures with different formation histories. Proper
motions for $\sim$$10^{7}$ stars, vs $\sim$$10^{6}$ with Roman, with uncertainties more than an order of magnitude
below the velocity dispersions of these components, separate them statistically and
allow a star formation history to be derived for each one. They will also reveal
kinematically distinct substructure left by past accretion events onto the NSC \citep[see][]{Feldmeier-Krause:2020pi}.
The same measurements tie the GC to the proper motion surveys of the bulge and disk by
Gaia and Roman, and give the crowding-free ground truth that Roman needs in the GC
(Sect.~\ref{sec:roman}).

\begin{table*}[!h]
\caption{Capabilities available after each epoch (upper rows, cumulative) and the epoch
at which each science question can be answered with the JWST data alone (lower rows).
\checkmark~answered, (\checkmark) first or partial constraints, $\times$ cannot be
addressed.}
\label{tab:epochs}
\begin{threeparttable}
\begin{tabular}{lcccc}
\hline
 & Epoch 1$^{a}$ & Epoch 2 & Epoch 3 & Epoch 4 \\
\hline
Years after epoch 1 & 0 & 4 & 8 & 12 \\
New NIRCam filters & F212N, F480M & F187N, F405N & F140M, F335M & F182M, F466N$^{b}$ \\
NIRCam bands in total & 2 & 4 & 6 & 8 \\
MIRI coverage & F770W (50\%) & F770W (100\%) & F2100W (50\%) & F2100W (100\%) \\
Proper motions$^{c}$ (mas\,yr$^{-1}$) & none & 0.125 & 0.063 & 0.042  \\
Accelerations$^{c}$ (mas\,yr$^{-2}$) & none & none & 0.08 (no redundancy) & 0.03  \\
Photometric epochs & 1 & 2 & 3 & 4 \\
\textbf{Q1} Missing young stars and clusters & (\checkmark) & (\checkmark) & \checkmark & \checkmark \\
\textbf{Q2} IMF & (\checkmark) & (\checkmark) & \checkmark & \checkmark \\
\textbf{Q3} Dust and ice & $\times$ & (\checkmark) & (\checkmark) & \checkmark \\
\textbf{Q4} 3D structure of the dense gas & (\checkmark) & (\checkmark) & \checkmark & \checkmark \\
\textbf{Q5} Present-day star formation rate & (\checkmark) & (\checkmark) & (\checkmark) & \checkmark \\
\textbf{Q6} Planet-forming disks & $\times$ & (\checkmark) & \checkmark & \checkmark \\
\textbf{Q7} Stellar-mass black holes & $\times$ & (\checkmark) & (\checkmark) & \checkmark \\
\textbf{Q8} Intermediate-mass black holes & $\times$ & $\times$ & (\checkmark) & \checkmark \\
\textbf{Q9} Assembly of the NSD and NSC & (\checkmark) & (\checkmark) & \checkmark & \checkmark \\
\textbf{Q10} Gravitational potential & $\times$ & (\checkmark) & \checkmark & \checkmark \\
\hline
\end{tabular}
\end{threeparttable}
\begin{tablenotes}[nosep]
\item $^{a}$JWST GC-Treasury
\item $^{b}$F430M (CO$_{2}$ ice) is an alternative to F466N (CO ice).
\item $^{c}$1$\sigma$ uncertainties for $\sigma = 0.5$\,mas per epoch (stars with $F212N \lesssim 18$) and four years between epochs.
\end{tablenotes}
\end{table*}

\subsection{Q10. What is the mass distribution in the inner 100\,pc of the Milky Way?}

The gravitational potential at the GC is key to interpreting the orbits of young clusters \citep{Hosek:2022om} and molecular clouds \citep[][see Q4]{Nogueras-Lara:2026ez}, and to constraining the mass-to-light ratio in this region, as well as the concentration of dark matter. Currently, the best constraints on the potential at these scales come from dynamical models fitted to the kinematics of only $\sim$$3000$ stars \citep{Sormani:2022dq,Vasiliev:2026sy}. These models determine the total NSD mass to $\sim$10\%, and leave uncertain whether the NSD is axisymmetric or barred. \citet{Fiteni:2026dk}  find hints of non-axisymmetry, which are however inconclusive because of the low-number statistics. Multi-epoch NIRCam astrometry will increase the number of stars by three orders of magnitude. This will make it possible to: (i) measure the enclosed mass  to $<1\%$ statistical precision, and map, for the first time, the transition between the NSC and the NSD; (ii) build dynamical models of  individual stellar populations and thus test the assumption made in the models so far that all populations in the NSD follow the same distribution function; (iii) detect or rule out a nuclear bar; (iv) potentially measure the gravitational potential directly through the statistically measured acceleration of millions of stars, although this may  require high angular resolution imaging observations beyond the mission lifetime of JWST, using future facilities (the expected acceleration of an individual star is roughly $2\times10^{-6}$\,mas\,yr$^{-2}$). 

\subsection{Summary}

Table~\ref{tab:epochs} shows how each epoch increases our ability to answer the aforementioned questions and give first constraints on the remaining ones. The required NIRCam and MIRI data will allow us to address the science questions through
\begin{itemize}[nosep]
\item Narrow-band multi-filter photometry to characterize the ionized interstellar medium as well as the sub-giant and main sequence stellar populations (Q2--Q6, Q9)
\item MIRI photometry to trace and characterize PAHs and YSOs (Q3, Q5, Q6)
\item Precision proper motions of roughly ten million stars (Q1, Q4, Q6, Q9, Q10)
\item Potentially measure statistically the acceleration of stars in the gravitational potential of the GC (Q10)
\item Accelerations of stars bound to high-mass stellar and intermediate-mass black holes (Q7, Q8)
\item Detect a large number of transients. Our preliminary analysis of the JWST Treasury Survey observations combined with other JWST observations of the GC  revealed  37 high-amplitude variables (confirmed by eye so far) in the 81 sq arcmin where there are multiple epochs. This indicates that we might find $\sim$ 650 high-amplitude (factor ten) variables and 900-2400 fast-moving stars ($>350$\,km\,s$^{-1}$).
\end{itemize}

The proposed multi-epoch observations cover the field of the GC Treasury Survey with similar instrumental settings and will therefore require a total time allocation of about 420\,h distributed over three epochs spaced three to five years apart.

\begin{figure}[!tb]
\includegraphics[width=\textwidth]{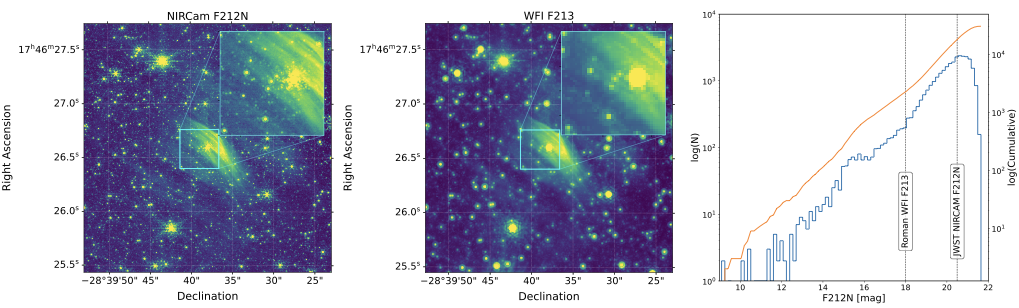}
\caption{Comparison of a GC field imaged with F212N NIRCam (GO 10678) and a simulation of this field using the Roman WFI F213 PSF and pixel scale. Right panel: Binned (blue) and cumulative (orange) luminosity function and approximate 50\% completeness limits for Roman \citep[approximated from HST WFC3 F153M][]{Hosek:2019vn} and JWST (Single exposure NRCA1 of o098 of JWST GC-Treasury - in combined exposures the completeness limit will improve to at least 21\,mag).\label{fig:comparison}}
\end{figure}

\section{Complementarity of JWST and Roman}
\label{sec:roman}

The multi-epoch JWST observations will be strongly complementary to the aforementioned Roman surveys. JWST will be able to reach below the main sequence turnoff, while Roman will be mostly limited to (sub-)giants in the GC.  JWST  will greatly enhance the Roman data by providing star-by-star extinction measurements, MIR photometry, much-improved depth and reduced systematic errors in the  highly crowded and extinguished GC, and identification of close/blended stars, whose photometry/astrometry can be significantly biased. The cumulative luminosity function from a single exposure with a single detector from JWST GC-Treasury compares an optimistically estimated completeness  limit of Roman WFI in this region with a conservative estimate for JWST NIRCam (based on a single one of six exposures). Even under these assumptions, NIRCam can detect more than five times as many stars as WFI (Fig.\,\ref{fig:comparison}). Finally, NIRCam and MIRI can penetrate the extremely extinguished regions of the GC.  Together Roman and JWST
herald a revolution in the interpretation of current/future data, bring together research in different sub-fields, and answer key open science questions with enormous legacy potential.

\section{A community resource}

Like the \href{https://sites.google.com/view/jwst-gc/home}{JWST GC Treasury Program}, the multi-epoch observations proposed in this White Paper will serve the entire community, from star and planet formation to galaxy evolution. The collaboration will be open for other researchers to join, as is already the case for GC-Treasury. The proposed program  will provide full field mosaics, cross-matched catalogs with JWST, Roman, and other observations, precision proper motions, and accelerations.

\bibliography{BibGC}

\end{document}